


\magnification 1200
\overfullrule 0 pt
\font\abs=cmr9
\font\ist=cmr8

\font\uit=cmu10

\def\CcC{{\hbox{\tenrm C\kern-.45em{\vrule height.67em width0.08em depth-.04em
\hskip.45em }}}}
\def\RrR{{\hbox{\tenrm I\kern-.17em{R}}}}
\def\HhH{{\hbox{\tenrm {I\kern-.18em{H}}\kern-.18em{I}}}}
\def\DdD{{\hbox{\tenrm {I\kern-.18em{D}}\kern-.36em {\vrule height.62em
width0.08em depth-.04em\hskip.36em}}}}
\def\ZzZ{{\hbox{\tenrm Z\kern-.31em{Z}}}}
\def\IiI{{\hbox{\tenrm I\kern-.19em{I}}}}
\def\NnN{{\hbox{\tenrm {I\kern-.18em{N}}\kern-.18em{I}}}}
\def\QqQ{{\hbox{\tenrm {{Q\kern-.54em{\vrule height.61em width0.05em
depth-.04em}\hskip.54em}\kern-.34em{\vrule height.59em width0.05em
depth-.04em}}
\hskip.34em}}}
\def\OoO{{\hbox{\tenrm {{O\kern-.54em{\vrule height.61em width0.05em
depth-.04em}\hskip.54em}\kern-.34em{\vrule height.59em width0.05em
depth-.04em}}
\hskip.34em}}}

\def\uq2{U_q({\uit su}(2))}

\def\fraz#1#2{{\strut\displaystyle #1\over\displaystyle #2}}

\def\part#1{\fraz{\partial}{\partial#1}}

\def\ii#1{\item{$\phantom{1}#1~$}}

\def\su2q{SU(2)_q}
\def\h1q{H(1)_q}

\def\nu{N_{1}}

\def\mapdown#1{\Big\| \rlap{$\vcenter{\hbox{$\scriptstyle#1$}}$}}
\def\mapcup{\bigcup}
\def\iinn{\rlap{{$\bigcup$}{$\!\!\! |$}}}
\def\mapin{\iinn}

\hsize= 15 truecm
\vsize= 22 truecm
\hoffset= 0.5 truecm
\voffset= 0 truecm

\null\vskip1.5truecm

\baselineskip= 13.75 pt
\footline={\hss\tenrm\folio\hss}
\centerline
{\bf  QUANTUM GROUPS, COHERENT STATES, SQUEEZING}
\smallskip
\centerline
{\bf AND LATTICE QUANTUM MECHANICS}
\bigskip
\centerline{
{\it    E.Celeghini ${}^1$, S.De Martino ${}^2$, S.De Siena ${}^2$,
             M.Rasetti ${}^3$ and G.Vitiello ${}^2$.}}
\bigskip
{\ist ${}^1$Dipartimento di Fisica - Universit\`a di Firenze and
INFN--Firenze, I50125 Firenze, Italy}
\footnote{}{\hskip -.85truecm {\abs E--mail: CELEGHINI@FI.INFN.IT\hfill
   DFF 188/9/93 Firenze}}

{\ist ${}^2$Dipartimento di Fisica - Universit\`a di Salerno and INFN--Napoli,
I84100 Salerno, Italy}

{\ist ${}^3$Dipartimento di Fisica and Unit\`a INFM -- Politecnico di
Torino, I10129 Torino, Italy}
\bigskip
\bigskip
\bigskip
\bigskip
\bigskip
\bigskip
\bigskip

\noindent{\bf Abstract.} {\abs By resorting to the Fock--Bargmann
representation, we incorporate the quantum Weyl--Heisenberg ($q$-WH) algebra
into the theory of entire analytic functions. The main tool is the realization
of the $q$--WH algebra in terms of finite difference operators. The physical
relevance of our study relies on the fact that coherent states (CS) are indeed
formulated in the space of entire analytic functions where they can be
rigorously expressed in terms of theta functions on the von Neumann lattice.
The r\^ole played by the finite difference operators and the relevance of the
lattice structure in the completeness of the CS system suggest that the
$q$--deformation of the WH algebra is an essential tool in the physics of
discretized (periodic) systems. In this latter context we define a quantum
mechanics formalism for lattice systems.}


\vfill\eject

\noindent {\bf 1. Introduction}
\bigskip
A great deal of attention and efforts have been devoted recently to the
mathematical structures referred to as $q$-groups$^{[1,2]}$,
newly discovered (or,
better, $\underline{\rm re}$disco\-vered, in that they are but specific
realizations of Hopf algebras) by theoretical physicists. These structures
promise to be very rich of physical meaning, and, although their skeleton
appears pretty well understood, there are however many properties which deserve
more study to be fully under control.

The interest in $q$-groups arose almost simultaneously in statistical mechanics
as well as in conformal theories, in solid state physics as in the study of
topologically non-trivial solutions to nonlinear equations. As a matter of
fact, the research in $q$-groups was indeed originated from physical problems.
One of the main interesting features resides in the $C^*$-algebraic properties
of the elements of their matricial representations, properties which can be
expected lead to fruitful connections with noncommutative geometries. Studies
in such directions are promising but at same time a comprehensive program is
far from being formulated.

A different situation occurs with the dual structures, the $q$-algebras (care
should be exerted here, as -- by an abuse of language now almost universally
adopted -- the name $q$-groups has been adopted to designate $q$-algebras as
well, even though they are in fact neither groups nor algebras in the customary
sense).

$q$-algebras are deformations of enveloping algebras of Lie algebras and, like
the latter, they have Hopf algebra features. The general properties of
$q$-algebras are much better known than those of $q$-groups, in particular for
the specific characteristics which relate them with concrete physical models.
Among the $q$-algebras, the non-semisimple ones are the less easy to handle, as
in fact it happens in general with non-semisimple structures. The
$q$-deformation of the Weyl-Heisenberg algebra ($q$-WH), as well as the WH
algebra, is not even a Hopf algebra; it has only the properties of a Hopf
superalgebra$^{[3]}$.

In view of the fundamental r\^ole of WH algebra in quantization, we believe
that its $q$-deformation deserves special attention. The WH algebra has in
quantum physics such powerful applications  as the harmonic oscillator,
coherent states, Jordan-Wigner realizations, and so on. $q$-WH algebra appears
to be equally useful in the study of the corresponding deformed structures.

In this paper we cast the study of the $q$-WH algebra in the frame of the
Fock-Bargmann representation (FBR) of Quantum Mechanics (QM). The reason for
this is that one of our objectives is to preserve, in our study of
$q$-deformations, the analytic structure of the corresponding Lie algebras and
therefore we need to operate in a scheme where analyticity is ensured.

As a first step we present a realization of the $q$-WH algebra in terms of
finite difference operators.
As a result we recognize that whenever a finite scale is
involved in a self-contained physical theory, then a $q$-deformation of the
algebra of dynamical observables occurs, with the $q$-parameter related with
the finite spacing, namely carrying the information about discreteness.
$q$-deformation is also expected in the presence of periodic conditions, since
periodicity is a special form of invariance under finite difference operators.

As a first conjecture, the physical content of the $q$-deformation thus emerges
in connection with discretized (periodic) systems.

The analytic properties of the FBR together with the von Neumann lattice
topological structure in the complex plane, make then manifest the relation
between the $q$-WH algebra and the coherent states (CS) in QM. Here, we obtain
a formal relation between the coherent state generator and the commutator of
$q$-WH creation and annihilation operators, which is thus recognized to be an
operator in the CS space. Moreover, theta functions, in terms of which CS are
expressed, also admit a representation in terms of $q$-deformed WH commutators
on the von Neumann lattice, allowing us to get further insight in the basic
unity of the various structures.

One additional successive step in the understanding of the physical meaning of
$q$-deformations is then achieved by realizing that the commutator of $q$-WH
creation and annihilation operators is, in the FBR, the squeezing generator for
CS, a result which confirms a conjecture, previously$^{[4]}$
formulated, whereby $q$-groups are the natural candidates to study
squeezed coherent states.

The relation established between CS and the $q$-WH algebra is of course of
great interest in view of the numerous interesting physical applications of the
CS formalism, and it may also open rich perspectives in Quantum Field Theory
(QFT) where the CS formalism is the key to study vacuum structure and boson
condensation.

The relevance of $q$-deformation to discretized system physics naturally leads
us to analyze the structure of Lattice Quantum Mechanics (LQM). We study it in
configuration space as well as in momentum space and show that LQM is
characterized in both cases by the algebra $E(2)$. The lattice CS, optimizing
the lattice position-momentum uncertainty relation, show that just $q$-WH is
the algebraic structure underlying the physics of lattice quantum systems. We
find that the commutator between $q$-WH creation and annihilation operators
acts as generator of the $U(1)$ subgroup of $E(2)$, giving rise to phase
variations in the complex plane. In this context, in the presence of a periodic
potential on the lattice, there emerges naturally a relation between $q$-WH
algebra and the Bloch functions, which further confirms the conjectured deeply
rooted presence of $q$-deformation within the dynamical structure of periodic
systems.

As a general remark, we should like to stress that it is only by fully
exploiting the FBR that we succeed in incorporating $q$-deformation of the WH
algebra into the theory of (entire) analytical functions. Such result may
deserve by itself further attention: in this way, indeed, it appears possible
to elucidate the deep r\^ole of $q$-WH algebra in the physics of lattice
quantum systems, coherent states and squeezing.

Through this paper we shall use units such that all relevant physical
quantities are dimensionless.

\bigskip
\noindent {\bf 2. $q$-Weyl-Heisenberg algebra,  Fock--Bargmann
representation and finite difference operators}
\bigskip
The quantum version of the Weyl-Heisenberg algebra
$$
[ a, a^\dagger ] = \IiI \quad ,\quad [ N, a ] = - a \quad ,\quad [ N,
a^\dagger ] = a^\dagger \quad , \eqno{(2.1)}
$$
is realized in terms of the set of operators $\{ a_q, {\bar a}_q, N_q ;
 ~q \in \CcC \}$,  with relations$^{\, [1,2]}$:
$$
[ N_q , a_q ] = - a_q \quad ,\quad [ N_q , {\bar a}_q ] = {\bar a}_q \quad
,\quad a_q {\bar a}_q - q^{-{1\over 2}} {\bar a}_q a_q = q^{{1\over 2}N_q}
\quad . \eqno(2.2)
$$

The structure lying behind (2.2) is a quantum superalgebra$^{\, [3]}$.
The notion of hermiticity associated with it has been studied in ref.
[4] in connection with the discussion of the squeezing of the
generalized coherent states $(GCS)_q$, defined in the usual Fock space
{}~${\cal K}$.

By introducing ~${\hat a}_q \equiv {\bar a}_q q^{N/2}$~, and setting
-- without loss of generality -- $N_q \equiv N$~, relations (2.2) can
be rewritten as
$$
[ N, a_q ] = - a_q \quad ,\quad [ N, {\hat a}_q ] = {\hat a}_q \quad , \quad
[ a_q, {\hat a}_q ] \equiv a_q {\hat a}_q - {\hat a}_q a_q = q^N  \quad .
\eqno(2.3)
$$

In the following we shall discuss the functional realization of eqs. (2.3) by
means of finite difference operators in the complex plane, in the Fock-Bargmann
representation of Quantum Mechanics$^{\, [5,6]}$.

Contrary to the usual coordinate or momentum representation of QM, where no
condition of analyticity is imposed upon the wavefunction, in the FBR any state
vector is described by an entire analytic function. The FBR operators, solution
of the commutation relations (2.1), are$^{\, [5,6]}$:
$$
N \to z {d\over dz} \quad ,\quad a^\dagger \to z \quad ,\quad a \to
{d\over dz} \quad .\eqno{(2.4)}
$$

The Hilbert space of entire analytic functions ${\cal F}$ has a well defined
inner product:
$$
<\psi _1 |\psi _2 > = \int {\bar \psi}_1(z) \psi _2(z) d\mu (z) \quad ,
\eqno{(2.5)}
$$
with suitable measure $d\mu$, where, in terms of the complete orthonormal set
of eigenkets $\{ |n> | n\in \NnN , |n> \in {\cal K} \}$ of $N$,
$\displaystyle{|\psi > \, = \sum_{n=0}^\infty c_n |n>\; , \; <\psi |\psi >\, =
\!\sum_{n=0}^\infty |c_n|^2}$ $= 1$.  One associates with $|\psi>$ the function
in ${\cal F}$
$$
\psi (z) = \sum_{n=0}^\infty c_n u_n(z) \quad , \quad u_n(z) = {z^n\over
\sqrt{n!}} \quad ,\quad (n~ \in {\NnN}_+) \; , \; u_0 (z) = 1 \quad .
\eqno(2.6)
$$

The set ~$\{ u_n(z)\}$~ provides an orthonormal basis in ~$\cal F$.
The r\^ole of the $\delta$-function is played in this representation,
assuming $d\mu (z)={1\over \pi} {\rm e}^{-|z|^2} d{\rm Re}z\,
d{\rm Im}z$, by
$$
\delta (z,z') = \sum_{n=0}^\infty u_n(z) {\bar u}_n(z') =
\exp(z z')\quad , \eqno{(2.7)}
$$
in that
$$
f(z) = \int \delta (z, z') f(z') d\mu (z') \quad . \eqno{(2.8)}
$$

Eqs. (2.6) provide thus the most general representation of
an entire analytic function in the $z$-plane. Note also that, from
eqs. (2.4) and (2.6),
$$\eqalign{a^\dagger ~u_n (z) ~~=~~ \sqrt{n + 1} ~~u_{n+1} (z) \; &, \;
a ~u_n (z) ~~=~~ \sqrt{n} ~~u_{n-1} (z) \; , \cr
N ~u_n (z) ~~=~~ a^\dagger a ~~u_n (z) ~~=&~~ z {d\over dz} ~~u_n (z) ~~=~~
n~ u_n (z) \; , \cr} \eqno{(2.9)}$$
as expected in view of the correspondence ~${\cal K}~ \to ~{\cal F}
{}~~( ~|n>~ \to u_n(z)~)$.

Let us now consider the finite difference operator ${\cal D}_q$
defined by:
$$
{\cal D}_q f(z) = {{f(q z) - f(z)}\over {(q-1) z}} \quad , \eqno{(2.10)}
$$
with ~$f(z) \in {\cal F}\; ,\; q = e^\zeta \; ,\; \zeta \in {\cal C}$ . ${\cal
D}_q$ is the so called $q$-derivative operator$^{\, [7]}$, which, for $q \to 1$
(i.e. $\zeta \to 0$), reduces to the standard derivative. By using the
representation (2.6) for $f$ and resorting to the last equality in eqs.(2.9),
it
may be written on ${\cal F}$ as
$$
{\cal D}_q = \bigl((q-1) z \bigr)^{-1} \Bigl( q^{z {d\over {dz}}} - 1\Bigr)
\quad . \eqno{(2.11)}
$$
Consistency between (2.10) and the above form of ${\cal D}_q$ can
be proven by first $''$normal ordering$''$
the operator ${\left ( z {d\over{dz}} \right )^n}$ in the form:
$$
\left ( z {d\over{dz}} \right )^n = \sum_{m=1}^n {\cal S}_n^{(m)} z^m
{{d^m}\over{dz^m}} \quad , \eqno{(2.12)}
$$
where ${\cal S}_n^{(m)}$ denotes the Stirling numbers of the second
kind defined by the recursion relations$^{\, [8]}$
$$
{\cal S}_{n+1}^{(m)} = m ~{\cal S}_n^{(m)} + {\cal S}_n^{(m-1)} \quad ,
\eqno{(2.13)}
$$
and then expanding in formal power series the exponential $\displaystyle{
\left ( q^{z {d\over {dz}}} - 1 \right )}$, keeping in mind the identity:
$$
{1\over{m!}} \left ( e^{\zeta} - 1 \right )^m = \sum_{n=m}^{\infty}
{\cal S}_n^{(m)} {{{\zeta}^n}\over{n!}} \quad . \eqno{(2.14)}
$$

${\cal D}_q$ generates, together with $z$ and ${d\over {dz}}$, the algebra:
$$
\bigl[ {\cal D}_q , z \bigr] = q^{z {d\over {dz}}} \quad ,\quad
\bigl[ z {d\over dz} , {\cal D}_q \bigr] = - {\cal D}_q \quad ,\quad
\bigl[ z {d\over dz} , z \bigr] = z \quad , \eqno{(2.15)}
$$
which can be recognized as a realization of relations (2.3) in the space
${\cal F}$. In the latter, operators ~$N$, $\hat a_q$ and $a_q$ can then
be associated with
$$
N \to z {d\over dz} \quad ,\quad {\hat a}_q \to z \quad ,\quad
a_q \to {\cal D}_q \quad , \eqno{(2.16)}
$$
with ~${\hat a}_q = {\hat a}_{q=1} = a^\dagger$~ and ~$\lim_{q\to1} a_q =
a$. The corresponding realization of (2.2) is
$$
N \to z {d\over dz} \quad ,\quad {\bar a}_q \to z q^{-z {d\over{dz}}} \quad ,
\quad a_q \to {\cal D}_q \quad . \eqno{(2.17)}
$$

There follows that the commutator  $[ a_q, {\hat a}_q ]$ acts in ${\cal F}$ as
$$
[ a_q, {\hat a}_q ]f(z) = q^{z{d\over dz}}f(z) = f(qz) \quad . \eqno{(2.18)}
$$

All of this suggests$^{\, [9]}$ that
whenever one deals with some finite scale ({\sl e.g.}
with some discrete structure, lattice or periodic system) which cannot be
reduced to the continuum by some limiting procedure, then a deformation of the
operator algebra acting in ${\cal F}$ should arise. Deformation of the operator
algebra is also expected whenever the system under study involves periodic
(analytic) functions, since periodicity is but a peculiar invariance under
finite difference operators.
\bigskip
\noindent {\bf 3. $q$-Weyl-Heisenberg algebra and coherent states}
\bigskip
In the following discussion of the connection of the $q$-WH algebra with the CS
formalism we shall confine ourselves to the case of a single complex variable
(one dimensional case). Extension to many complex variables would proceed along
the same lines as the extension of customary CS to several degrees of
freedom$^{\, [6]}$ and it is straightforward. We begin by observing that it is
just by exploiting the theory of entire analytic functions that the
Fock--Bargmann representation provides a simple and transparent frame to study
the usual CS$^{\, [6,10]}$. Actually, writing the latter in
the form
$$
|\alpha > = {\cal D}(\alpha ) |0> ~~~;\quad
a |\alpha> = \alpha |\alpha>\quad , \quad a |0> = 0\quad , \quad
\alpha \in \CcC  \quad , \eqno{(3.1)}
$$
$$
|\alpha> = \exp\biggl({-|\alpha|^2\over 2} \biggr) \sum_{n=0}^\infty {{\alpha
^n}\over {\sqrt{n!}}} |n> = \exp\biggl({-|\alpha|^2\over 2}\biggr)
\sum_{n=0}^\infty u_n(\alpha) |n> \quad , \eqno{(3.2)}
$$
the relation between the CS and the entire analytic function basis ~$\{
u_n(z) \}$ is easily made explicit: $u_n (z) = {\rm e}^{{1\over 2}|z|^2}
<n|z>$.  The unitary displacement operator ${\cal D}(\alpha)$ in (3.1) is
given by:
$$
\eqalignno{
{\cal D}(\alpha) = \exp\bigl(&\alpha a^\dagger -{\bar \alpha} a \bigr)
= \exp\biggl(-{{|\alpha|^2}\over 2}\biggr) \exp\bigl(\alpha a^\dagger\bigr)
\exp\bigl(-{\bar \alpha} ~a\bigr)  \; , &(3.3a) \cr
&= \exp\biggl({{|\alpha|^2}\over 2}\biggr) \exp\bigl(-{\bar \alpha}
a\bigr)\exp\bigl(\alpha a^\dagger\bigr) \; , &(3.3b)\cr }
$$
and the following relations hold
$$
{\cal D}(\alpha) {\cal D}(\beta) = \exp\bigl( i {\it Im}(\alpha {\bar
\beta})\bigr) {\cal D}(\alpha + \beta) \quad , \eqno{(3.4)}
$$
$$ {\cal D}(\alpha) {\cal D}(\beta) = \exp\bigl( 2 i {\it Im}(\alpha {\bar
\beta})\bigr) {\cal D}(\beta) {\cal D}(\alpha) \quad . \eqno{(3.5)}
$$

Eq. (3.5) is the well known integral representation of the Heisenberg
commutation relations (2.1), also called the Weyl integral representation (see
e.g. ref. [6]).

In order to relate the $q$-deformed Heisenberg algebra (2.3) (or (2.2)) to the
CS generator ${\cal D}(\alpha)$, we observe that, upon rewriting eq. (3.4)
with $\alpha = (q-1) z$ and $\beta = z$, one gets
$$
\eqalignno{
<n|{\cal D}\bigl( &(q-1) z\bigr) |z> =
<n| \exp\Bigl( i {\it Im}(q-1)|z|^2\Bigr) |q z> = &(3.6a) \cr
&= \exp\biggl({{q-{\bar q}}\over 2} |z|^2\biggr)
\exp \biggl(-{{|q z|^2}\over 2}\biggr) u_n(q z) \quad .  &(3.6b) \cr}
$$
On the other hand we have, still from (3.2),
$$
<n| q^N |z> = \exp\biggl(-{{|z|^2}\over 2}\biggr) u_n(q z)\quad .
\eqno{(3.7)}
$$
Thus, from (2.3), (3.6b) and (3.7),
$$
<n| [ a_q, {\hat a}_q ] |z> = \exp\biggl(-(1-{\bar q})(1+q) {{|z|^2}\over
2}\biggr) <n| {\cal D}\bigl((q-1) z\bigr) |z> \quad . \eqno{(3.8)}
$$

Use of (3.6a) gives
$$
<n| [ a_q, {\hat a}_q ] |z> = \exp\biggl((|q|^2-1)
{{|z|^2}\over 2}\biggr) <n|q z> \quad , \eqno{(3.9)}
$$
for any $|n>$ (and for any $|z>$). Thus we can write
$$
\exp\biggl((1-|q|^2){{|z|^2}\over 2}\biggr)  [ a_q, {\hat a}_q ] |z> =
|q z> \quad . \eqno{(3.10)}
$$
Finally, use of eq. (3.3a) leads to
$$
{\cal D}(\alpha) f(z) = \exp\biggl(-{{|\alpha|^2}\over 2}\biggr) \exp(\alpha z)
f(z - {\bar \alpha}) \quad ,\quad f \in {\cal F} \quad . \eqno{(3.11)}
$$

By setting ~$\alpha = (1-{\bar q}) {\bar z}$~ in eq. (3.11),
we have then, in view of eq. (2.18),
$$
[ a_q, {\hat a}_q ] f(z) = \exp\biggl(-(1-{\bar q})(1+q)
{{|z|^2}\over 2}\biggr) {\cal D}\bigl((1-{\bar q}) {\bar z}\bigr) f(z)
\quad . \eqno{(3.12)}
$$

These results establish the relation of the quantum algebra (2.3) (or (2.2))
with the theory of CS (eqs. (3.8)$\div$(3.10)) and with the theory of
entire analytic functions (eqs. (2.18), (3.7) and (3.12)).

It is interesting to observe that the commutator $[ a_q, {\hat a}_q ]$  acts as
shift operator from ~$|z>$~ to ~$|q z>$~ (notice that the exponential factor in
eq. (3.10) simply generates the correct normalization); it acts as the
$z$-dilatation operator ~$(z \to q z)$~ in the space of entire analytic
functions and it acts as the $U(1)$ generator of phase variations in
the $z$-plane when $q = e^\zeta$, with $\zeta$ pure imaginary, $\zeta = i
\theta \,:  \, z ~\to e^{i \theta} z$.

We also observe that eqs. (3.8), (3.10) and (3.12) provide a non linear
realization of the quantum algebra (2.3) in terms of $a$ and $a^\dagger$.
Conversely, the nonlinear operator ${\cal D}(\alpha)$ is represented by the
linear form ~$[ a_q, {\hat a}_q ]$ in the algebra.

We finally recall that in order to extract a complete set of CS $\{
|z_n>\}$~$\bigl (\, \{ z_n\}$ a discrete set of points in the $z$-plane minus
the origin $\, \bigr )$ from the overcomplete set $\{ |z>\}$ it is necessary to
introduce a regular lattice $L$ $^{\, [5,6]}$ . By closely following the
procedure of ref. [6] we recall that the set $\{ |z_n>\}$ (with exclusion of
the vacuum state $|0>$) is complete if the lattice elementary cell has area $S
= \pi$ $^{\, [6,10]}$ ($L$ is called, in this case, the von Neumann lattice).
The points (or lattice vectors) $z_n$ are given by $z_n = \sum_i \mu _i \omega
_i\, ,\, i=1,2$ with $\mu_i \in \ZzZ$.  The lattice periods $\omega _i$ are
assumed to be linearly independent, {\sl i.e.} ${\rm Im}\, \omega _1 {\bar
\omega _2} \not= 0$. The proof of completeness is established by invoking
square integrability along with analyticity$^{\, [11]}$.

The possibility of extracting the complete set $\{ |z_n>\}$ makes more
significant the above presented relation between $q$-algebras and CS, and once
more we stress the central r\^ole of the underlying discrete lattice structure.

\bigskip
\noindent {\bf 4. ~ $q$-Weyl-Heisenberg and theta functions}
\smallskip
The lattice structure is of crucial relevance in the relation
between the theta functions and the complete system of CS.

A deep interrelation, essentially amounting to equivalence, is known to exist
among three different ways of viewing {\sl theta functions}:
\item{a)} as classical holomorphic functions in (several) complex variables
${\bf z}$ and the period (matrix) $\tau$;
\item{b)} as matrix coefficients of a representation of the Heisenberg and
metaplectic group;
\item{c)} as sections of line bundles on abelian varieties or the moduli
space of abelian varieties.

\noindent Such an equivalence is thoroughly analyzed by Mumford in ref.~[12],
where its mathematical structure is explored in full detail. Here it is our aim
to show how these profound mathematical results are related to some physically
interesting features. For the sake of completeness, we simply summarize
hereafter the basic mathematical concepts on which our analysis will be based,
reporting them in the concise $''${\sl flow chart}$''$ in $Tab. I$.
In such
chart ${\cal H}$ denotes the Hilbert space (${\cal H}^*$ its dual) of
representations of the Heisenberg group $H$. ~$H$ acts unitarily on the
elements of ${\cal H}$,
which are holomorphic functions, by
$$
U_{(A,B)} f(Z) = \exp{\left \{ i \pi {\rm Tr} \left [ A^t (TAQ + B)\right ]
\right \}}\, f(Z+TAQ+B) \quad . \eqno{(4.1)}
$$
$A$, $B$, $Z$, $T$, and $Q$ are complex matrices and in $Tab. I$
$\iota$ means {\sl
isomorphism}, ${\cal H}_{\pm \infty}$ are the completions of ${\cal H}$ to the
space of holomorphic functions such that $f(X)= {\cal O}\bigl ( |\! | X|\!
|^{\pm n} e^{{\pi\over 2} \Xi (X,X)}\bigr )$, for some integer $n$, with
quadratic (negative if ${\rm Im}T$ is negative) weight $\Xi (X,X) = X^t ({\rm
Im}T)^{-1} X$, $e_{\bf z}$ $= \sum_{N\in \ZzZ} e_* (N/2) \delta_N$, (where
$e_*(N/2)\in \{\pm 1\}$ is a quadratic form on ${1\over 2}\ZzZ / \ZzZ $),
$\sigma$ denotes the cocycle of the group action, $W_T \equiv \{ {\rm
span~of}\, f \mapsto X\cdot f\}$, and $f_T$ is the (highest weight) element in
the Heisenberg representation, {\sl unique} up to a scalar, which is
annihilated by $W_T$.

$$
{\cal H}^* \otimes {\cal H}
$$
$$
\mapdown{\iota}
$$
$$
\left [ \matrix{{\rm All~functions}\, f \, {\rm on}\, \RrR^{\otimes 2}  \cr
                {\rm on~which}\, H\otimes H \, {\rm acts}\cr} \right ]
$$
$$
\mapdown{}
$$
$$
\left [ \matrix{{\rm Space~spanned~by~functions} \cr
                <U_{(A,B)}f , g> \, , \, f\in {\cal H}_{\infty}\, , \,
                g\in {\cal H}_{-\infty} \cr }\right ]
$$
$$
\mapcup \qquad \qquad \qquad \qquad \mapcup
$$
$$
W_T \, {\rm right-annihilator} \qquad \qquad \sigma\left ( \ZzZ^{\otimes 2}
\right )\! -\!{\rm left-invariants}
$$
$$
\mapdown{} \qquad \qquad \qquad \qquad \qquad \qquad \qquad \qquad \mapdown{}
$$
$$
\left [ \matrix{{\rm Space~spanned~by~functions} \cr
                <U_{(A,B)}f_T , f> \, , \, f\in {\cal H}_{-\infty}\cr }\right ]
\qquad \qquad
\left [ \matrix{{\rm Space~spanned~by~functions} \cr
                <U_{(A,B)}f , e_{\bf z}> \, , \, f\in {-\cal H}_{\infty}
                \cr }\right ]
$$
$$
\mapdown{} \qquad \qquad \qquad \qquad \qquad \qquad \qquad \qquad \mapdown{}
$$
$$
e^{- {\pi\over 2} \Xi} \, \left [ \matrix{{\rm Fock~space}\cr
                                          {\cal F}(\CcC ,T)_{- \infty}
                                          \cr }\right ]
\qquad \qquad
\left [ \matrix{{\rm Quasi-periodic~space}\cr
                l^2(\RrR^{\otimes 2}/\!\ZzZ^{\otimes 2}\cr }\right ]
$$
$$
\mapdown{\iota} \qquad \qquad \qquad \qquad \mapdown{\iota}
$$
$$
{\cal H}^* \qquad \qquad \qquad \qquad {\!\!{\cal H}}
$$
$$
{\!\!\!\!\mapin}\; \qquad \qquad \, \qquad \qquad \mapin
$$
$$
\left [ \matrix{{\rm The~unique~function}\cr
                \vartheta^Q \left [ {A\atop B} \right ] (T) =
                < U_{(A,B)} f_T , e_{\bf z}> \cr} \right ]
$$

\bigskip

\centerline{$Tab. I.$}

In general,
$$
\eqalign{
\vartheta^Q &\left [ {A\atop B} \right ] (T) \cr
&= \sum_{N\in \ZzZ} \exp{\left \{ i \pi {\rm Tr}\, \left [ (N+A)^t T (N+A) Q
+2\, (N+A)^t (Z+B) \right ]\right \}} \quad . \cr } \eqno{(4.2)}
$$

We shall relate, in what follows, the general scheme of the
$''${\sl flow chart}$''$  with the
properties of coherent states, in a quite elementary way, which, indeed, is
nothing but a rephrasing of the flow chart depicted above.

In order to do so, we look for the common eigenvectors $|\theta>$ of the CS
operators ${\cal D}(z_n)$ associated to the regular lattice $L$ $^{\, [6]}$. A
common set of eigenvectors exists only if all the ${\cal D}(z_n)$ commute,
which happens, of course, when the ${\cal D}(\omega _i)$ commute. From eq.
(3.4) we see that ${\cal D}(\omega _i), i=1,2$, commute when  ${\it Im} \omega
_1 {\bar \omega _2} =  k \pi $ with $k$ integer, namely just when $L$ is a Von
Neumann lattice. Without loss of generality, we set $k = 1$, which is
sufficient to guarantee that operators  ${\cal D}(z_n)$ generate a complete set
of CS.

The states $|\theta>$ are characterized by two real numbers $\epsilon _1$ and
$\epsilon _2$, $|\theta> \equiv |\theta _\epsilon>$, since, as required, they
are eigenvectors of ${\cal D}(\omega _i)$:
$$
{\cal D}(\omega _i) |\theta_\epsilon> = e^{i \pi \epsilon _i}
|\theta_\epsilon>\quad , \quad i=1,2 \quad , \quad 0 \leq \epsilon _i \leq 2
\quad . \eqno{(4.3)}
$$
The vector $|\theta _\epsilon>$ corresponds to a point on the two-dimensional
torus. $|\theta _\epsilon>$ belongs to a space which is the extension$^{\,
[6]}$ of the Hilbert space where the operators ${\cal D}(z)$ act.  The action
of ${\cal D}(z)$ on $ |\theta _\epsilon>$ generates a set of generalized
coherent states $ |\theta _z>$. The system of CS  $|\theta _\epsilon>$ in the
FBR may be associated with an entire analytic function, say $\theta (z)$,
$|\theta _\epsilon> \to \theta (z)$. Use of eqs. (4.3) and (3.4) gives
$$
{\cal D}(z_m) |\theta_\epsilon> = e^{i \pi F_\epsilon(m)} |\theta _\epsilon>
\quad , \eqno{(4.4)}
$$
with $z_m = m_1 \omega _1 + m_2 \omega _2$ an arbitrary lattice vector and
$$
F_\epsilon(m) = m_1 m_2 + m_1 \epsilon _1 + m_2 \epsilon _2 \quad .
\eqno{(4.5)}
$$

Moreover, eq. (3.11) with $\bar \alpha ~= -z_m$ shows that eq. (4.4) may be
written  as
$$
\theta_\epsilon (z+z_m) = \exp\bigl({i \pi F_\epsilon(-m)}\bigr) \exp
\biggl({{|z|^2}\over 2}\biggr) \exp({\bar z_m} z) \theta_\epsilon(z) \quad ,
\eqno{(4.6)}
$$
which is the functional equation for the theta functions$^{\, [6,13]}$ (see eq.
(4.2)). The strict relation between the theta functions and the CS system is
thus established. We emphasize that such relation is obtained by considering
the CS system corresponding to the admissible lattice $L$.

A solution of (4.6) can be expressed as follows$^{\, [6]}$:
$$
\eqalignno{
\theta_\epsilon(z) = &\sum_m e^{- i \pi F_\epsilon(m)}
{\cal D}(-{\bar z}_m) f(z) \quad   &(4.7a) \cr
= \sum_m e^{- i \pi F_\epsilon(m)} &\exp\biggl(-{{|z_m|^2}\over 2}\biggr)
   \exp\bigl(-{\bar z}_m z\bigr) f(z) \quad , &(4.7b) \cr }
$$
where $f(z)$ is an arbitrary entire function such that the series (4.7b) is
convergent.

We now have all the ingredients necessary to establish the relation
between $q$-WH algebra and theta functions.
Let us make first explicit the dependence of the deformation parameter $q$ on
the lattice periods $\omega _i$ by writing ~$q = q_m =e^{\zeta_m}$~, with
$\zeta _m$ a vector on the von Neumann lattice $L$.

By setting $z_m = (q_m -1) z$, from eqs. (4.7b) and (3.11), (3.12) we obtain
$$
\theta_\epsilon(q_m z) = [ a_{q_m}, {\hat a_{q_m}}] \theta_\epsilon(z)
\quad , \eqno{(4.8)}
$$
which, by use of (4.6), gives
$$
[ a_{q_m}, {\hat a_{q_m}}] \theta_\epsilon(z) = \exp\bigl({i \pi
F_\epsilon(-m)}\bigr) \exp\biggl(-(1-{\bar q}_m)(1+q_m) {{|z|^2}\over 2}\biggr)
\theta_\epsilon(z) \quad . \eqno{(4.9)}
$$

Moreover, use of (3.12) in eq. (4.7a) gives
$$
\theta_\epsilon(z) = \sum_m  \exp\bigl({- i \pi F_\epsilon(m)}\bigr)
\exp\biggl((1-{\bar q}_m)(1+q_m)~{{|z|^2}\over 2}\biggr)
[ a_{q_m}, {\hat a_{q_m}}]\, f(z) \quad .  \eqno{(4.10)}
$$

Eqs. (4.8)$\div$(4.10) exhibit the relation between the $q$-algebra (2.3) and
the theta functions: eq. (4.8) is the functional equation for the theta
function corresponding to (4.2), (4.6) and, together with (4.9), shows the
explicit action of  $[ a_q, {\hat a}_q ]$ on
$\theta_\epsilon(z)$. Eq. (4.10) expresses the solution of (4.6) and of (4.8)
(i.e. the theta function) in terms of the application of the operator $ [
a_{q_m}, {\hat a_{q_m}}]$ to the arbitrary entire function $f(z)$.

The natural way to exhibit the topological as well as the dynamical
meaning of this representation is to recall that theta functions are
actually the quantum propagator for systems whose space of dynamical
variables $\cal M$, transitive under some compact Lie group $\cal G$, is
multiply connected$^{\, [14]}$.
In particular, in the simple case in which ~${\cal G} \sim SO(2)$, ~$\cal M$ is
just the 1-torus (circle) $T^{(1)}$. Assuming ${\cal G} \sim SO(2)\otimes
SO(2)$~, ~due to the composition properties of theta functions, or,
equivalently, of the CS displacement operators (eq.(3.4)) would lead
just to ${\cal M} \sim T^{(2)} \sim T^{(1)}\times T^{(1)}$, the 2-torus
CS-state manifold of the $\theta_\epsilon$'s$^{\, [12]}$.
In the former case (extension to the latter is immediate) the free
dynamics, for which the orbits are geodesics and hence the lagrangian
${\cal L}$ is the manifold metrics, leads
-- by path integration in the universal covering space of $\cal M$ --
to the propagator
$$
\eqalignno{
K(\varphi_1 ,t_1 | \varphi_2 ,&t_2 ) = \int {\cal D}[\varphi (t)]
e^{i S[\varphi (t)]/\hbar} \Rightarrow \cr
\Rightarrow K_{\mu}(\phi ,\tau ) &= {1\over{2\pi}} \exp{\left (i \left [{{\mu
\phi}\over {2\pi}} - {{\mu^2}\over{8 \pi^2 \Theta}}\right ]\right )}
\, \vartheta_3 \left ( {1\over 2} \phi - {{\mu}\over{8\pi \Theta}};
- {1\over{2\pi \Theta}} \right ) ~~ ,~~
&(4.11) \cr }
$$
where $S[\varphi ]$ denotes the action integral
$S[\varphi (t)] = \int_{t_{in}}^{t_{fin}} {\cal L}[\varphi (t)] dt \; ,$
$\Theta$ is a dimensionless parameter depending on the system
(e.g. the moment of inertia for a rigid rotator) and on the process
duration $\tau = t_{fin} - t_{in}$; whereas $\phi =
\varphi_{fin} - \varphi_{in}$
is the corresponding variation of the variable parametrizing the system
state (e.g. the polar angle defining the position on $T^{(1)}$).
$\mu$ is the phase coherence parameter whereby the global rotation $\cal R$:
$\varphi_{fin} \to \varphi_{fin} +
2 \pi$ implies ${\cal R}(K_\mu) = e^{i \mu} K_\mu$.
$\theta_3(z;s)$ is the standard Jacobi theta function$^{\, [13]}$
$$
\vartheta_3 (z;s) = \sum_{n = - \infty}^{+ \infty} e^{i \pi s n^2}
e^{2 i n z} \quad .
\eqno{(4.12)}
$$

Eq. (4.11) shows how the representative functions $\theta_{\epsilon}$ carry
indeed dynamical as well as topological information, in that their structure is
non-trivial only in view of the homotopical non-triviality of the
configuration space.

It is finally interesting to realize the action of $[ a_{q_m}, {\hat a_{q_m}}]$
on a conformal image ${\tilde {\cal F}}$ of ${\cal F}$ in terms of the
variables $u$ and ${d\over {du}}$ defined by
$$
u = \log z \quad , \quad {d\over {du}} = z {d\over {dz}} \quad . \eqno{(4.13)}
$$

These have canonical commutator $[ u, {d\over {du}}] = \IiI$ in
${\tilde {\cal F}}$. We have (cfr. (2.18)),
by setting $q_m = e^{\zeta _m}$ and
$f(z) = f(\exp(\log z)) = {\tilde f}(u)$,
$$
[ a_{q_m}, {\hat a_{q_m}}] f(z) = {q_m}^{z {d\over {dz}}} f(z) =
\exp\bigl(\zeta_m {d\over {du}}\bigr) {\tilde f}(u) = {\tilde f}(u+\zeta_m)
= f(q_m z) \quad , \eqno{(4.14)}
$$
i.e. the commutator $[ a_{q_m}, {\hat a_{q_m}}]$ generates a translation of $u$
by the lattice vector $\zeta_m$. Thus, in the $u$, ${d\over {du}}$ variables,
the relation of the $q$-algebra (2.3) with the CS generator ${\cal
D}(\alpha_m)$, which acts indeed as a ``translation" operator on the lattice L
is even more transparent. Also, the functional equation (4.8) for theta
functions looks more familiar when written in terms of the $u$ variable in
${\tilde {\cal F}}$:
$$
{\tilde \theta}_\epsilon(u+\zeta_m) = [ a_{q_m}, {\hat a_{q_m}}]  {\tilde
\theta}_\epsilon(u) \quad . \eqno{(4.15)}
$$

These results lead us once more to conclude that the existence of a quantum
deformed dynamical algebra signals the presence of a lattice length in the
theory and provides the natural framework for the physics of discretized
systems.
\bigskip
\noindent {\bf 5. $q$-Weyl-Heisenberg algebra and the squeezing generator}
\bigskip
Let us consider the harmonic oscillator Hamiltonian
$$H ~=~ {1\over 2} \bigl({\hat p}_z^2 + {\hat z}^2\bigr) \; ,
\eqno{(5.1)}$$
\noindent
where ${\hat p}_z = - i {d\over {dz}}$ and $[ {\hat z}, {\hat p}_z] = i$,
over a Hilbert space of states identified with the space of entire
analytic functions ~${\cal F}$.  Introduce, as customary, the operators
$$
{\alpha} = {1\over {\sqrt{2}}} \bigl( {\hat z} + i {\hat p}_z
\bigr) \quad ,\quad { \alpha}^\dagger = {1\over {\sqrt{2}}} \bigl( {\hat z} -
i {\hat p}_z\bigr)\quad ,\quad [\alpha, \alpha^\dagger ] = \IiI \quad ,
\eqno{(5.2)}
$$
in terms of which
$$
H = {\alpha}^\dagger {\alpha} + {1\over 2}\quad . \eqno{(5.3)}
$$

The ground state is described by the Gaussian wavefunction $\psi^0(z) = (\pi
)^{-{1\over 4}}$ $\exp\bigl(-{{z^2}\over {2}}\bigr) \;$. It is immediate to
observe that, via the definition (5.2), we have
$$
2 z {d\over {dz}} f(z) = \bigl[\bigl({\alpha}^2 - {\alpha}^{\dagger 2}\bigr) -
1\bigr] f(z) \quad ,\quad f \in {\cal F} \quad  , \eqno{(5.4)}
$$
and the operator  $[ a_q, {\hat a}_q]$ acts as the squeezing operator
${\hat {\cal S}}(\zeta)$ $^{\, [15]}$ on any state $\psi (z)$ (see also eq.
(2.18)):
$$
\eqalign{
[ a_q, {\hat a}_q] \psi (z) = \exp\Bigl(\zeta &z {d\over dz}\Bigr) \psi (z) =
{1\over{\sqrt q}} \exp\Bigl({\zeta\over 2}\bigl(\alpha^2 -
{\alpha^\dagger}^2\bigr) \Bigr)\psi (z)\cr
&= {1\over{\sqrt q}} {\hat {\cal S}}(\zeta) \psi (z) =
{1\over{\sqrt q}} \psi_{s}(z)\quad , \cr}
\eqno{(5.5)}
$$
with $q = e^\zeta$ and $\psi _s(z)$ denoting the squeezed state. In the case of
the ground state ${\psi^0}_s(z) = \psi ^0(q z) = \bigl({q^2\over
{\pi}}\bigr)^{1\over 4} \exp\bigl(-{{q^2 z^2}\over {2}}\bigr)\;$.
The minimum Heisenberg uncertainty relation holds for  $\psi^0(z)$
as well as for  ${\psi^0}_s(z)$ and $\psi_s (z)$:
$$
\Delta z \Delta p_z = {1\over 2} \quad , \eqno{(5.6)}
$$
with $\Delta z = {\sqrt{1\over 2}} = \Delta p_z$ for  $\psi^0(z)$ and $\Delta
{z} = {1\over q} {\sqrt{1\over 2}}, \Delta { p}_z = q{\sqrt{1\over 2}}$ for
$\psi_s (z)$ (and of course ${\psi^0}_s$).

We thus conclude that the $q$-deformation parameter plays the r\^ole of
squeezing parameter and the commutator $[ a_q, {\hat a}_q ]$ is, up to a
numerical factor, the squeezing generator with respect to the operators
$\alpha$ and $\alpha^\dagger$ :
$$
[ a_q, {\hat a}_q] = {1\over{\sqrt q}}~\exp\biggl({\zeta \over 2}\bigl(\alpha^2
- {\alpha^\dagger}^2\bigr)\biggr)\quad . \eqno{(5.7)}
$$

On the other hand, it is well known$^{\, [15]}$ that the right hand side of
(5.7) is an $SU(1,1)$ group element. In fact by defining $K_{-} = {1\over
2}\alpha ^2$, $K_{+} = {1\over 2}\alpha^{\dagger 2}$, $K_{z} = {1\over
2}(\alpha^\dagger \alpha + {1\over 2}) = {1\over 2}H$, one easily checks they
close the algebra $su(1,1)$.

Let us finally observe that in view of the holomorphy conditions on $f(z)$
($z = x + iy$)
$$
{d\over dz} f(z) = {d\over dx} f(z) = -i {d\over dy} f(z) \quad ,\quad
f \in {\cal F} \quad ,\eqno{(5.8)}
$$
one finds
$$
z {d\over dz} = z {d\over dx} = x {d\over dx} + iy{d\over dx} =
-ix {d\over dy} + iy{d\over dx} = x p_y - y p_x \equiv L \quad , \eqno{(5.9)}
$$
with $L$ an angular momentum operator. For $\zeta = i \theta$, with $\theta$
real, the commutator $[ a_q, {\hat a}_q ]$ acts then in ${\cal F}$ as the
$U(1)$ group element
$$[ a_q, {\hat a}_q ] = e^{i \theta L} \quad , \eqno{(5.10)}
$$
as already observed above.

\bigskip
\noindent {\bf 6. Lattice  Quantum Mechanics.}
\bigskip

\noindent In this section we recall first the structure of
Lattice Quantum Mechanics
(LQM) in order to relate it to $q$-WH
and construct lattice CS minimizing the lattice
position-momentum uncertainty relation.

We limit ourselves to 1-dimensional lattice (extension to more dimensions is
straightforward). 1-dimensional lattice quantum system is defined
on the configurational Hilbert space ${\cal G} = l^2(\epsilon \ZzZ )$
where $\ZzZ$ denotes
the set of integers $n$ and $\epsilon$ is the lattice spacing.

The hermitian position operator, ${\hat x}_{\epsilon}$, is
defined as
$$
{\hat x}_{\epsilon} f(x_{n}) = x_{n} f(x_{n}) = \epsilon n f(x_{n})
\quad , \quad f \in {\cal G} \quad,  \eqno{(6.1)}
$$
whereas the hermitian lattice momentum operator ${\hat p}_{\epsilon}$ is
defined by
$$
{\hat p}_{\epsilon} f(x_{n}) = -i {\cal D}_{\epsilon} f(x_{n})\quad ,
\eqno (6.2)
$$
where ${\cal D}_{\epsilon}$ is the symmetrized, finite difference
gradient
$$
{\cal D}_{\epsilon} f(x_{n}) = (2 \epsilon)^{-1}
[f(x_{n+1}) - f(x_{n-1})]
\quad . \eqno (6.3)
$$
The dual momentum space representation of the above operators is
therefore
$$
{\hat x}_{\epsilon} f(k) = i {d\over {dk}} f(k)\quad , \eqno (6.4a)
$$
$$
{\hat p}_{\epsilon} f(k) = \epsilon^{-2} {\sin (k \epsilon)}
f(k) \quad , \eqno (6.4b)
$$
respectively, where $f(k)$ is the Fourier conjugate of $f(x_n)$, $k$ belonging
to the first Brillouin zone (BZ), $|k| \leq \pi/\epsilon$.

Over ${\cal G}$ we have the following commutation relation between
position and momentum
$$
[{\hat x}_{\epsilon} , {\hat p}_{\epsilon} ] f(x_{n}) =
{i\over 2} \left (f(x_{n+1}) + f(x_{n-1})\right ) \quad . \eqno (6.5)
$$

The operators ${\hat P}_\epsilon = \cos (\epsilon  {\hat p} )\; , \; {\hat p} =
-i {d\over {dx}}$, and ${\hat x}_{\epsilon}$ generate the algebra $E(2)$:
$$
[{\hat x}_{\epsilon} , {\hat p}_{\epsilon} ] = i {\hat P} ~~,~~ [{\hat
x}_{\epsilon} , {\hat P} ] = -i \epsilon^2 {\hat p}_{\epsilon} ~~,~~ [{\hat
P} , {\hat p}_{\epsilon} ] = 0  \quad .  \eqno (6.6)
$$
Relations (6.6) follow of course from the
simple observation that
$$
f(x_{n \pm 1}) \equiv f(x_{n} \pm \epsilon) =
\bigl(\exp(\pm i \epsilon {\hat p} \bigr) f(x_{n}) \quad ,
\eqno (6.7)
$$
provided the derivatives of $f(x_{n})$ exist at any order.

Let us notice that the above  $E(2)$ algebra contracts to the Weyl--Heisenberg
algebra in the limit $\epsilon \rightarrow 0$:
the discrete lattice spacing $\epsilon$ plays thus the r\^ole of
deformation parameter.

Upon introducing, in momentum space, the angle $\phi = k \epsilon$,
$-\pi \leq \phi \leq \pi$, on the
unit circle, we set
$$
L_3 = -i {d\over {d\phi}} = - {i\over \epsilon} {d\over {dk}} \quad ,\quad
L_1 = \cos \phi \quad ,\quad L_2 = \sin \phi \quad .  \eqno{(6.8)}
$$
The algebra (6.6) takes then the usual form of the (angular momentum)
$E(2)$ algebra:
$$
[ L_3, L_1 ] = i L_2 \; ,\;  [ L_3, L_2 ] = -i L_1 \; ,\;  [ L_1, L_2 ] = 0 \;
. \eqno{(6.9)}
$$
Following standard procedures$^{\, [17,18]}$, one gets the uncertainty
inequalities
$$
\eqalign{
\Delta^{\!2} \left({\hat x}_{\epsilon} \right)
\Delta^{\!2}  \left({\hat p}_{\epsilon} \right) &\geq
{1 \over 4} {\langle \cos(k \epsilon) \rangle}^{2} \quad ,\cr
\Delta^{\!2} \left({\hat x}_{\epsilon}\right)
 \Delta^{\!2} \left(\cos(k \epsilon) \right) &\geq
{1 \over 4}{\epsilon^{2}} {\langle {\sin(k \epsilon)}
\rangle}^{2} \quad , \cr} \eqno (6.10)
$$
where $\langle {\hat A} \rangle = \int dk \Psi^{*}(k) {\hat A} \Psi (k)\;  $
denotes quantum expectation on the lattice and $ \Delta^{\!2}({\hat A}) =
\langle {\hat A}^2 \rangle - {\langle {\hat A} \rangle}^2 $ the (square)
variance. The continuum limit $\epsilon \rightarrow 0$ corresponds to opening
the circle into a line. In $d$ dimensions the limit  $\epsilon \rightarrow 0$,
is an isometric and conformal mapping of the torus on the plane
(decompactification)$^{\, [16]}$.
The states minimizing the uncertainty products of
eqs.(6.10) must satisfy$^{[17,18]}$
$$
\left({\hat x}_{\epsilon} + i \gamma {\hat p}_{\epsilon} \right) \Psi = \lambda
\Psi \quad , \eqno (6.11)
$$
where $\lambda = \langle {\hat x}_{\epsilon} \rangle + i \gamma \langle {\hat
p}_{\epsilon} \rangle$ , and $\gamma$ is connected with the variances
$\Delta\left({\hat x}_{\epsilon} \right)$~,~$
\Delta\left({\hat p}_{\epsilon} \right)$
of position and momentum.
Relation (6.11) becomes, in momentum space,
$$
\left[{d \over{d(\epsilon k)}} + {\bar \gamma} \sin (\epsilon k)\right]
\Psi(k) = -i {\bar \lambda} \Psi(k)\quad , \eqno (6.12)
$$
where ${\bar \lambda} = \lambda \epsilon^{-1} \,, \, {\bar \gamma} = \gamma
\epsilon^{-2}$. Its solution is
$$
\Psi(k) = G \exp \left [{\bar \gamma} \cos (\epsilon k) - i {\bar \lambda}
\epsilon k \right ]\quad , \eqno (6.13)
$$
where the normalization constant $G$ is given by $G = {{2\pi}\over \epsilon}
I_{0}(2{\bar \gamma})$ , $I_{0}$ denoting the modified Bessel function of the
first kind of order $0$. Notice that, in the continuum limit $\epsilon \to 0$,
the Fourier transform ${\tilde \Psi}(x)$ of (6.13) becomes
$$
\tilde\Psi (x) = \left({\gamma \pi}\right)^{-{1\over 4}}\, \exp \left\{- \left
[ (2\gamma)^{-1} (x - \langle {\hat x} \rangle)^2 + i \langle {\hat p} \rangle
(x - \langle {\hat x} \rangle)\right ] \right\} \quad , \eqno{(6.14)}
$$
which is just the minimum uncertainty wavefunction given by Schr\"odinger$^{\,
[19]}$. By setting $z = <x> + i \gamma <p>$, eq. (6.14) defines the usual
coherent states$^{\, [6]}$. The wave-function (6.13) is the coherent state for
a system with discretized position and momentum or, equivalently,
endowed with some periodic
constraint.

\bigskip
\noindent{\bf 7. $q$-Weyl-Heisenberg algebra and lattice quantum mechanics}
\smallskip
We aim now to show that also the structure of LQM is underlined by $q$-WH.
In order to do so, we
introduce a conformal image $\tilde {\cal F}$ of the configurational Hilbert
space ${\cal G}$ by defining $z = e^{ix}$, in terms of which
$$
i {\hat p} f(x_{n}) = {d \over {dx}} f(x_{n}) = i z {d \over {dz}} {\tilde
f}(z_{n})\quad ,\quad z_{n} = e^{ix_{n}}\quad ,\quad {\tilde f} \in
{\tilde {\cal F}}\quad . \eqno (7.1)
$$
The assumption that the derivatives of $f(x_n)$ of any order exist (cf.eq
(6.7)) implies of course that $\tilde f(z_n)$ is an analytic function, which
we also assume to be entire, on the unit circle ${\cal C}$.
$z_n = e^{ix_{n}}$ is of course
a discrete set of points in the $z$-plane belonging to ${\cal C}$. The function
$\tilde f(z_n)$ may thus be represented by means of the expansion in the basis
$u_m(z_n)$ as in eq. (2.6).

Recalling that
$$
\left [{\hat p} , e^{ix} \right ] = e^{ix}\;,\; \left [{\hat p} , e^{-ix}{\hat
p} \right ] = - e^{-ix}{\hat p}\quad ,\quad \left [e^{-ix}{\hat p} , e^{ix}
\right ] = 1\quad ,\eqno{(7.2)}
$$
provides a realization of the WH algebra (2.1) on the unit circle, suggests
that the realization (2.4) of the FBR may be adopted in $\tilde {\cal F}$ for
the LQM as well. Eqs. (6.3), (6.5) and (6.7) are now written, respectively,
as
$$
i {{\hat p}_{\epsilon}} f(x_{n}) =
{\cal D}_{\epsilon} f(x_{n}) =
i\, {{q^N - q^{-N}} \over {2 \log \, q}} {\tilde f}(z_{n}) =
i\, {\epsilon}^{-1} {\sin(\epsilon N)} {\tilde f}(z_{n})\quad ,
\eqno (7.3)
$$
$$
[{\hat x}_{\epsilon} , {\hat p}_{\epsilon}]f(x_{n}) =
i \cos(\epsilon N) {\tilde f}(z_{n})\quad , \eqno (7.4)
$$
and
$$
f(x_{n+1}) = e^{i \epsilon {\hat p}} f(x_{n}) = {q^N} {\tilde f}(z_{n}) =
{\tilde f}(q z_{n}) = {\tilde f}(z_{n+1})\quad , \eqno (7.5)
$$
where we have set $q =e^{i \epsilon}$, and $q^N$ was introduced through the
$q$-WH algebra (2.3).

This makes clear that the algebraic structure of LQM is intimately related with
the $q$-WH algebra, the $q$-deformation parameter being determined by the
discrete lattice length $\epsilon = -i \log \, q$.

The same conclusion can be reached constructing LQM in momentum space.
In such a case one may consider a conformal image $\tilde {\cal H}$ of
the Hilbert space in the momentum representation
by setting  $ z = e^{i \phi}\;$ , where we introduced once more
the angular variable over ${\cal C}$~,~$ \phi = k \epsilon
\; ,\; -\pi \leq \phi \leq \pi \; $, so that (cf. eq.(6.8))
$$
-i {d \over {d \phi}} =
-{i\over \epsilon} {d\over {dk}} =
z {d \over {dz}} \quad .\eqno (7.6)
$$

The functions belonging to $\tilde {\cal H}$ are assumed to be entire analytic
functions for which expansions of the form (2.6) hold. On the other hand,
$$
L_{3} f(\phi) = -i {d \over {d \phi}} f(\phi) = z {d \over {dz}} {\tilde f}(z)
= N {\tilde f}(z) \quad ,\quad {\tilde f} \in {\tilde {\cal H}} \quad , \eqno
(7.7)
$$
and
$$
f(\phi+\epsilon) = e^{i \epsilon L_{3}} f(\phi) =
q^N {\tilde f}(z) = {\tilde f} (qz) \quad . \eqno (7.8)
$$
The $E(2)$ algebra (6.9) is now realized in the momentum FBR as
$$
[L_{1} , L_{3}] {\tilde f}(z) = -i L_{2} {\tilde f}(z)\; , \;
[L_{2} , L_{3}] {\tilde f}(z) = i L_{1} {\tilde f}(z)\; , \;
[L_{1} , L_{2}] {\tilde f}(z) = 0 \; , \eqno (7.9)
$$
with ${\tilde f} \in {\tilde  {\cal H}}$, and the identifications
(see eq.(6.8))
$$
L_{1} ={ {z+{\bar z}} \over 2}\; ,\; L_{2} = {{z-{\bar z}} \over {2i}}\; ,
\; L_{3} = z {d \over {dz}}\; ,\; L_{+} = z\; , \; L_{-} = {\bar z}\; .
\eqno (7.10)$$

One can check that in this representation
$[a_{q}, {\hat a}_{q}]$ is nothing but the $e^{i \epsilon
L_{3}}$ group  element of $E(2)$, which recovers eq. (5.12). We  also  note
that $z^{n} = e^{i n \phi}$, $n$ integer, is the eigenfunction of $L_{3}$
associated with the integer eigenvalue $n$ of the  number operator in the FBR
$$
L_{3} z^{n} = N z^{n} = n z^{n}\quad . \eqno (7.11)
$$

The  functions $z = e^{i \phi}$ and $z^{n}$ play also a crucial r\^ole in the
Bloch functions theory$^{\, [20]}$. Suppose we  have a periodic potential
$V(x_{n}) = V(x_{n} + \epsilon)$ on the  lattice. Bloch  theorem then ensures
the existence of solutions of the related Schr\"odinger equation of the form
$$
\psi(x_{n}) = e^{\pm i k x_{n}} v_{k}(x_{n})\quad , \eqno (7.12)
$$
with $v_{k}(x_{n}) = v_{k}(x_{n} + \epsilon)$. $\psi(x_{n})$ is the Bloch
function.

Let us limit ourself for simplicity to considering the plus sign in (7.12).
$\psi(x_{n})$ has the property
$$
\psi(x_{n} + \epsilon) = e^{i k \epsilon} \psi(x_{n}) =
z \psi(x_{n})\quad . \eqno (7.13)
$$
We  thus see that the choice of the variable $z= e^{i k \epsilon}$ turns out
to be natural in the case  of periodic potentials:
$$
\psi(x_{n}) = z^{n} v_{k}(x_{n})\quad ,\quad
\psi(x_{n} + \epsilon) = z^{n+1} v_{k}(x_{n})\quad . \eqno (7.14)
$$
Since $z^{n} = (z_{n})^k$, from eq. (7.5)
$$
q^N (z_{n})^k = (q z_{n})^k = (e^{i \epsilon} e^{i x_{n}})^k =
e^{i k \epsilon (n+1)} = z^{n+1}\quad , \eqno (7.15)
$$
where $q^N$ is understood as defined on $\tilde {\cal F}$.
We  thus have in $\tilde {\cal F}$
$$
\psi(x_{n} + \epsilon) = [a_{q} , {\hat a}_{q}] (z_{n})^k u_{k}(x_{n}) =
[a_{q} , {\hat a}_{q}] \psi(x_{n})\quad , \eqno (7.16)
$$
namely the condition implementing the Bloch function periodicity
features is realized by the same operator in $q$-WH algebra acting,
in the FBR, as dilatation (see eq.(2.18) and compare eq. (7.13) with
(7.14)-(7.16)). This shows that Bloch functions provide in fact a
representation of the $q$-WH algebra.
\bigskip
\noindent {\bf 8. Conclusions}
\bigskip

In this paper we have considered the $q$-deformation of the WH algebra  in
connection with typical problems in QM: discretized systems, coherent states,
squeezing, lattice quantum mechanics, finite difference operators, periodic
systems. These have been proposed as physically relevant examples
where $q$-deformation play a r\^ole. Such a collection of applications
is not only
interesting on its own, but also as a {\it laboratory} where problems
characterized by the common feature of a discrete structure are treated in the
framework of (entire) analytic functions theory.

The underlying philosophy  is that  whenever one deals with a quantum system
defined on a countable set of degrees of freedom, then one has to work in the
space of the related analytical functions and the structure of the operator
algebra is the $q$-deformed one. In this sense, the word {\it deformation} does
not sound as the most appropriate, since $q$-algebras provide an algorithm of
wide physical application, and not an exception for {\it deformed} situations.
Such a general philosophy appears to find very concrete
support in the results discussed
in this paper.

{}From the point of view of group theory , we used the well known  mapping of a
$q$-algebra into the universal  enveloping algebra of the corresponding Lie
structure; to be specific, the mapping of finite difference operators into
functions of differential operators, which can be indeed achieved only by
operating on $C^\infty$ functions. This was
the main reason to work with FBR and
to introduce the conformal image of both configuration and momentum
space in the study of LQM.

An interesting and natural development, namely the extension from QM to QFT by
considering the infinite volume limit of the lattice system, is at present
under study.  We conjecture that the unitarily inequivalent representations of
the canonical commutation relations are parametrized by the $q$-deformation
parameter in such a limit. Different values of the lattice spacing are thus
described by inequivalent representations. In this framework  finite
temperature and dissipative systems$^{\,[21]}$
may find an appropriate unified
description.

We gratefully aknowledge useful discussions with O.V.Man'ko, Y.Katriel
and A.I.Solomon.

\vfill\eject
\baselineskip= 16 pt
\nopagenumbers

\phantom{xxxxx}
\bigskip

\line{{\bf References.}\hfill}
\bigskip

\ii 1   L.C Biedenharn, J. Phys. {\bf A22} (1989) 117

\ii 2   A.J. Macfarlane, J. Phys. {\bf A22} (1989) 4581

\ii 3   E.Celeghini, T.D.Palev and M.Tarlini, Mod. Phys. Lett {\bf B5}
        (1991)187

\ii 4   E. Celeghini, M. Rasetti and G. Vitiello, Phys. Rev. Lett. {\bf 66}
        (1991) 2056

\ii 5   V.A. Fock, Z. Phys. {\bf 49} (1928) 339

	V. Bargmann, Comm. Pure and Appl. Math. {\bf 14} (1961) 187

\ii 6   A. Perelemov, {\it Generalized Coherent States and
	Their Applications},  Springer-Verlag
	Berlin  Heidelberg  1986

\ii 7   L.C. Biedenharn, and M.A. Lohe , Comm. Math. Phys. {\bf 146},
          483 (1992)

\ii {\phantom{7}} F.H. Jackson, Messenger Math. {\bf 38}, 57 (1909)

\ii {\phantom{7}} T.H. Koornwinder,
        Nederl. Acad. Wetensch. Proc. Ser. {\bf A92},
        97 (1989)

\ii {\phantom{7}} D. I. Fivel, J. Phys. {\bf A24}, 3575 (1991)

\ii 8   M. Abramowitz and I.A. Stegun, {\it Handbook of Mathematical
Functions},
        Dover Publications, Inc., New York, 1972

\ii 9   E. Celeghini, S. De Martino, S. De Siena, M. Rasetti and G. Vitiello,
        Mod. Phys. Lett. {\bf B}, in press

\ii {10}   J.R. Klauder and B. Skagerstam, {\it Coherent States},
        World Scientific 1985

\ii {11}  V. Bargmann, P. Butera, L. Girardello and J.R. Klauder, Rep. on Math.
        Phys. {\bf 2} (1971) 221

\ii {12}  D. Mumford, {\it Tata Lectures on Theta, III}, Birkh\"auser Boston,
        Inc., 1991

\ii {13}  R. Bellman, {\it A brief introduction to theta functions},
          Holt, Rinehart and Winston, N.Y. 1961

\ii {14}  L. Schulman, Phys. Rev. {\bf 176} (1968) 1558

\ii {15}  H.P. Yuen, Phys. Rev. {\bf A13} (1976) 2226

\ii {\phantom{15}} E. Celeghini, M. Rasetti, M. Tarlini and G. Vitiello, Mod.
        Phys. Lett. {\bf B3} (1989) 1213

\ii {16}  C. De Concini and G. Vitiello, Nucl. Phys. {\bf B116} (1976) 141

\ii {17}  R. Jackiw, J. Math. Phys. {\bf 9}, (1968) 339

\ii {18}  P. Carruthers and M. Nieto, Rev. Mod. Phys. {\bf 40} (1968) 411

\ii {19}  E. Schr\"odinger, Naturwissenschaften {\bf 14} (1926) 664

\ii {20}  A.J. Dekker, {\it Solid State Physics}, Prentice-Hall, Englewood
          Cliffs 1957

\ii {21}  E. Celeghini, M. Rasetti and G. Vitiello, Annals of Phys.(N.Y.)
          {\bf 215} (1992) 156

\vfill\eject
\bye